\documentclass[12pt]{iopart}
\usepackage{epsf}
\begin{document}
\title{Quantum channel based on correlated twin laser beams}
\author{Constantin V. Usenko\dag\ and Vladyslav C. Usenko\ddag
\footnote[3]{To whom correspondence should be addressed (usenko@univ.kiev.ua)} }

\address{\dag\ National Shevchenko University of Kyiv, Department of Theoretical Physics, Kyiv, Ukraine}

\address{\ddag\ Insitute of Physics of National Academy of Science, Kyiv, Ukraine}

\begin{abstract}
This work is the development and analysis of the recently proposed quantum cryptographic protocol,
based on the use of the two-mode coherently correlated states. The protocol is supplied with the
cyrptographic control procedures. The channel error properties and stability against eavesdropping 
are examined. State detection features are proposed.
\end{abstract}

\section{Introduction.}

The goals of quantum cryptography and secure quantum communications 
\cite{qc1, qc2, qc3} can be achieved using various protocols, which were 
developed and realized \cite{entprot, fourexp} in the past years on the basis 
of the quantum entanglement \cite{ent1, entprot} 
of weak beams and the single \cite{single1, single2} or few photon states \cite{four}, 
mostly by means of adjusting and detecting their polarization angles \cite{polar}. 

Another method, based on the usage of the two-mode coherently correlated (TMCC) beams
was proposed recently \cite{tmcc, tmcc2}. In this case the secure cryptographic key is generated
by the laser shot noise and dublicated through the quantum channel. Unlike the single or few photon schemes, 
which require large numbers of transmission reiterations to obtain the statistically significant 
results, the TMCC beam can be intensive enough to make each single measurement 
statistically significant and thus to use single impulse for each 
piece of information, and remain cryptographically steady.

In this work we analyse the error and security properties of the quantum channels, based on the
TMCC-beams and propose some additions to the TMCC-based cryptographic protocol.

The two-mode coherently correlated state is the way we refer to the 
generalized coherent state \cite{per}, which can be described by its presentation through series by Fock states:

\begin{equation}
\label{eq:tmcc}
\left| \lambda \right\rangle =\frac{1}{\sqrt{I_0\left(2\left|\lambda\right|\right)}}\sum_{n=0}^\infty {\frac{\lambda ^n}{n!}\left| {nn} 
\right\rangle } 
\end{equation}

Here we use the designation $\left| {nn} \right\rangle = \left| n 
\right\rangle _1 \otimes \left| n \right\rangle _2 $, where $\left| n 
\right\rangle _1 $ and $\left| n \right\rangle _2 $stand for the states of 
the $1^{st}$ and $2^{nd}$ mode accordingly, represented by their photon 
numbers. The states (\ref{eq:tmcc}) are not the eigenstates for each of the annihilation operators 
separately, but are the eigenstates for the product of annihilation 
operators: 
\begin{equation}
	a_1 a_2 \left| \lambda \right\rangle = \lambda \left| \lambda 
	\right\rangle . 
\end{equation}
Such states can also be obtained from the zero state:

\begin{equation}
\label{eq:tmccground}
\left| \lambda \right\rangle = \frac{1}{\sqrt{I_0\left(2\left|\lambda\right|\right)}}I_0 (\lambda a_1^ + a_2^ + )\left| 0 
\right\rangle 
\end{equation}

In this work we assume that two laser beams, which are 
propagating independently from each other, correspond to the two modes of 
the TMCC state. States of beams are mutually correlated. (Surely, the TMCC 
state can also be represented in another way, for example, as a beam 
consisting of two correlated polarizations \cite{similar1})

\newpage

\section{Beam measurement}
Unlike the usual non-correlated coherent states, which
show their quasiclassical properties in the fact, that the mean value of a vector-potential of
a corresponding beam is not equal to 0, the mean value of a vector-potential of a TMCC-beam 
and any other characteristic, which is linear in field, turns to be equal to 0, because during 
the averaging by the 
1$^{st}$ mode the $a_1$ converts $\left| {n,n} \right\rangle $ to $\left| 
{n - 1,n} \right\rangle $, which is orthogonal to all the present state 
terms, so $\left\langle {\lambda _i } \right|a_i \left| {\lambda _i } 
\right\rangle = 0$, and so the quasiclassical properties in their usual meaning are absent in 
this case.

The intensity of the beam's radiation, registered by an observer is proportional to the 
mean of the $N = a^ + a$ operator, which is the number of the photons 
in the corresponding mode. 
The mean observable values, which characterize the results of the 
measurements of the beam are:

\begin{equation}
\label{eq:meann}
\left\langle {N } \right\rangle = \left\langle \lambda \right|a^ 
+ a \left| \lambda \right\rangle , \left\langle {N^2 } \right\rangle = 
\left\langle \lambda \right|a^ + a a^ + a \left| 
\lambda \right\rangle 
\end{equation}

These characteristics are squared in field, and thus their mean values don't turn to 
zero (surely, this fact is not specific for the TMCC-states, because the usual non-correlated 
states and processes, like the heat propagation, show the same properties).

Assuming the state expression (\ref{eq:tmcc}) we obtain

\begin{equation}
\label{eq:meann2}
\langle {N } \rangle = \frac{1}{I_0 ( 2| \lambda | )}\sum^\infty_{n=0} {n \frac{| \lambda |^{2n}}{n!^2}}
 , \langle {N^2 } \rangle = \frac{1}{I_0 ( 2| \lambda | )}\sum^\infty_{n=0} {n^2 \frac{| \lambda |^{2n}}{n!^2}}
\end{equation}

The mean number of registered photons is

\begin{equation}
\langle N \rangle = \sum_{n=0}^\infty nP_n (\lambda)
\end{equation}

The probability of registering n photons depends on the intensity of a beam:

\begin{equation}
\label{eq:nphotprob}
P_n(\lambda) = \frac{|\lambda|^{2n}}{I_0 (2|\lambda|)n!^2} 
\end{equation}

\begin{figure}[htbp]
\begin{center}
	\epsfbox{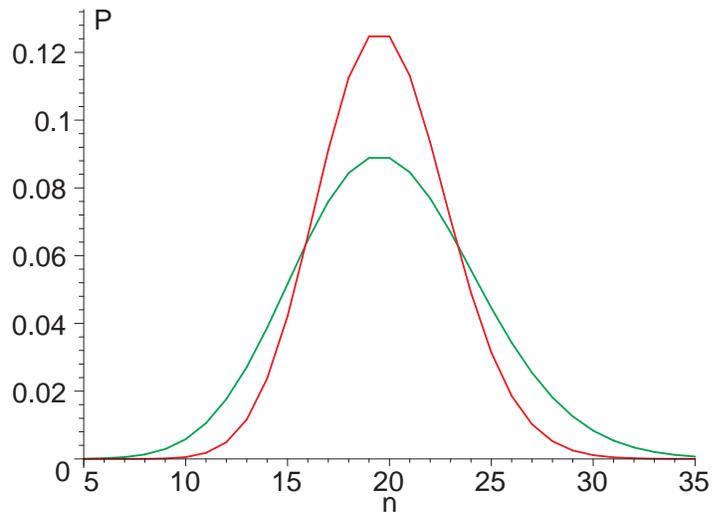}
	\caption{ Probability of different photon numbers registration distribution for the TMCC-beam (green) and the corresponding distribution for a usual coherent beam (red)}
	\label{plot_pn}
\end{center}
\end{figure}

An important feature of this distribution 
is the quick (proportional to $n!^2 $) decreasing dependence of the registration probability
on the photon number. This circumstance makes the experimental identification of the
TMCC-states quite convenient. The distribution of the probability of different photon 
numbers registration along with the corresponding distribution for a usual coherent beam 
are given at the (\fref{plot_pn}).

Taking into account (\ref{eq:nphotprob}) the expressions for the mean and mean 
square values of the registered photon numbers (\ref{eq:meann2}) turn to:

\begin{equation}
\label{eq:meann3}
\langle N \rangle = \frac{|\lambda|^2 I_1(2|\lambda|)}{I_0 (2|\lambda|)} , 
\left\langle {N^2 } \right\rangle = \left| \lambda \right|^2
\end{equation}

The type of the beam statistics can by characterized by the
Mandel parameter:

\begin{equation}
\label{eq:mandel}
Q=\frac{<N^2>-<N>^2}{<N>}-1
\end{equation}

\begin{figure}[htbp]
\begin{center}
	\epsfbox{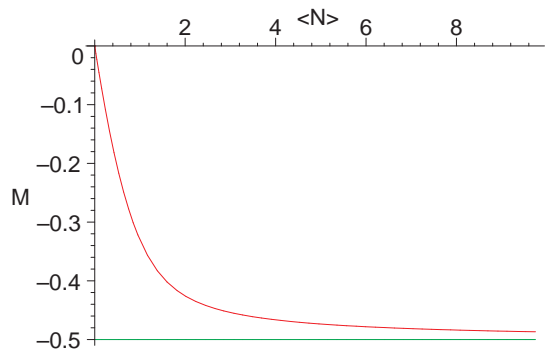}
	\caption{ Mandel parameter dependence on the mean photon numbers value for a TMCC beam}
	\label{plot_mandel}
\end{center}
\end{figure}

The dependece of the Mandel parameter (\ref{eq:mandel}) on the mean photon number is
given at the (\fref{plot_mandel}). One can easily see that the TMCC beam shows evident
sub-Poisson statistics even at small intensities.

The measurements have the statistical uncertainty, caused by quantum 
fluctuations. This uncertainty can be characterized 
by the corresponding dispersion:

\begin{equation}
\sigma ^2 = \langle {N^2} \rangle - \langle 
{N} \rangle ^2
\end{equation}

Taking into account (\ref{eq:meann3}) we get the following 
expression:

\begin{equation}
\label{eq:sigma}
\sigma ^2 = \left| \lambda \right|^2\left( {1 - \left( {\frac{I_1 
\left( {2\left| \lambda \right|} \right)}{I_0 \left( {2\left| \lambda 
\right|} \right)}} \right)^2} \right)
\end{equation}

\begin{figure}[htbp]
\begin{center}
	\epsfbox{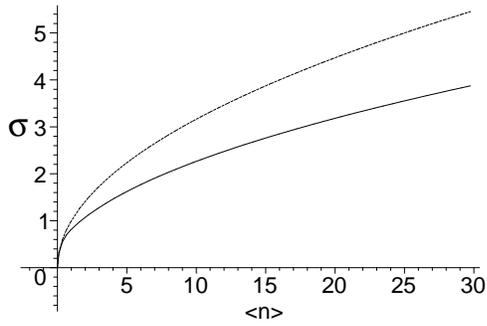}
	\caption{The dependencies of the measurement results dispersion on the mean photon number for the TMCC-beam (solid line) and a usual coherent beam (dotted line)}
	\label{plot_dispers}
\end{center}
\end{figure}

The dependencies of the measurement results dispersion on the mean photon number for
the TMCC-beam and a usual correlated beam are given at (\fref{plot_dispers}). 
One can see that there are significant differences for the TMCC and the Poisson beam distributions. 
This fact can be used for the TMCC-states identification.

\newpage

\section{Communication via quantum channel}

\begin{figure}[htbp]
	\epsfbox{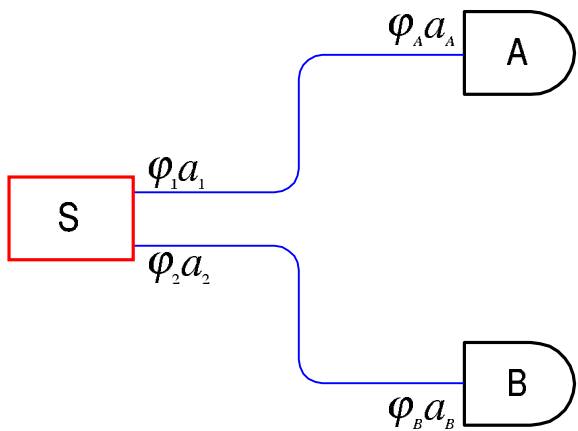}
	\caption{Quantum channel between two parties with a TMCC source}
	\label{scheme1}
\end{figure}

Let we have to establish a secure quantum channel between two parties (\fref{scheme1}). Alice has the laser on her side, which produces two beams in the TMCC 
state. The optical channel is organized in such a way, that Alice receives one 
of the modes, the first, for example, i.e. $\varphi _A \equiv \varphi _1 
$,$\varphi _A (x_A ,t_0 ) = 1$ , and Bob receives another one, i.e. $\varphi _B 
\equiv \varphi _2 $ ,$\varphi _B (x_B ,t_0 ) = 1$ at any moment of 
measurement $t_0 $, where $x_A $and $x_B $are Alice's and Bob's locations 
respectively. Accordingly, Alice cannot measure the Bob's beam and vice 
versa:$\varphi _B (x_A ,t_0 ) = 0$, $\varphi _A (x_B ,t_0 ) = 0$. At that 
the vector-potential of the TMCC-beam is:

\begin{equation}
A = \varphi _A^\ast (x,t)a_A^ + + \varphi _A (x,t)a_A + \varphi _B^\ast 
(x,t)a_B^ + + \varphi _B (x,t)a_B 
\end{equation}

As it was noticed above, the quasiclassical properties in their usual meaning are absent in 
the case of a TMCC-beam, but they become apparent in the spatial correlation function,
which characterizes the interdependence of the results of measurements 
taken by Alice and Bob:

\begin{equation}
g_{AB} = < N_A N_B > - < N_A > < N_B > 
\end{equation}

It's useful to describe the channel quality by the relative correlation, 
which is

\begin{equation}
\rho _{AB} = \frac{ < N_A N_B > - < N_A > < N_B > }{\sigma _A \sigma _B }
\end{equation}

The main feature of the TMCC state is that the value $\rho _{AB} $ is 
exactly equal to 1, while in the case of non-correlated beams we would get 
$\rho _{AB} = 0$. This means that the measurements of the photon numbers, 
received by Alice and Bob, each with her/his own detector, not only show the same 
mean values, but even have the same deflection from the mean values.

The laser beam is the semi-classical radiation with well defined phase, but 
due to the uncertainty principle for the number of photons and the 
phase of the radiation, there is a large enough uncertainty in the photon 
numbers, this can be seen from the dispersion expression (\ref{eq:sigma}). Thus one can 
observe the noise, which is similar to the shot noise in an electron tube. 
In the TMCC radiation the characteristics of such noise for each of the 
modes are amazingly well correlated to each other. This fact 
enables the use of such radiation for generation of a random code, which 
will be equally good received by two mutually remote detectors.

\subsection{The protocol}

The following scheme can be used for the TMCC-based protocol. The laser is 
set up to produce the constant mean number of photons during the session and 
both parties know this number. At some moment Alice and Bob start the 
measurements. They detect the number of photons at unit time  by measuring 
the integrated intensity of the corresponding incoming beam. If the number 
of photons for the specific unit time is larger than the known expected mean 
(which is due to the shot noise), the next bit of the generated code is 
considered to have the value ``1''. If the measured number is less than 
the expected mean, the next bit is considered to be equal to ``0'':
 
\begin{equation}
\label{protocol}
B ={\left\{
\begin{array}{l}
\{n \leq [<N>]\} \rightarrow 0 \\ \{n > [<N>]\} \rightarrow 1
\end{array}\right.}
\end{equation}

Upon the receipt of a sufficient number of bits (the code), both Bob and Alice 
divide them in half, each obtaining two bit sequences (half-codes). Bob encodes one 
sequence with another, using the XOR (excluding OR) logical operation, 
and sends this encoded half-code to Alice using any public channel. 
Alice uses any of her half-codes to
decode the code she has got from Bob using the same XOR operation. She compares
the result of this operation with another of her half-codes. If
all the bits coincide, this means that Alice and Bob both have the same
code, which can be used as a cryptographic key for encoding their communication.
Otherwise they have to repeat the key generation and transfer procedure and check 
the channel for the possible eavedropping if the procedure fails again.

\subsection{Quantum channel error analysis}

Let the parties of the secret key transmission procedure are using the protocol 
described above, thus they estimate the value of the next bit by comparing the 
actual registered photon number to the average. The probability of detecting "0" 
bit value then is

\begin{equation}
P_{(0)} (\lambda) = \sum_{n=0}^{[\langle N \rangle]}P_n(\lambda)
\end{equation}

The noise is present in the channel and it may increase the number of the registered 
photons. We suppose that the noise is thermal and assume that 
it may, with some probability, cause an appearance of one and no more than one 
additional photon in any of the modes during the time of a bit detection. 
We will denote the probability of a noise photon detection as $\epsilon$ 
and refer to it as the noise factor.

We suppose that the channel is qualitative enough to transfer the impulse at the 
required distance without losing any single photon, thus errors are possible only 
due to the appearance of the noise photons.

An error, when Alice registers "0" bit value and Bob registers "1" may occur upon 
the joint realization of two events. The first is that Alice detects the maximum 
number of possible for the "0" bit value, which is, according to the proposed protocol, 
equal to the integer part of $\langle n \rangle$. The second is that in addition to this number Bob 
detects the appearance of a noise photon. The opposite situation, when Alice gets the 
"1" bit value and Bob registers "0" is possible when the noise photon was detected 
by Alice, the probability of such error is the same. 

The probability of the realization of a state, which consists of the maximum possible 
for the "0" bit value photons and, at the same time, is detected as "0", is the 
relation between the corresponding probabilities:

\begin{equation}
P_{max(0)} = \frac{P_{[\langle N \rangle]}}{P_{(0)}} = \frac{P_{[\langle N \rangle]}}{\sum_{n=0}^{[\langle N \rangle]}P_n(\lambda)}
\end{equation}

We will refer to this probability as to the error factor.
So the probability of an error during the bit registration is equal to the product 
of the noise and error factors:

\begin{equation}
P_{err(0)}(\lambda)=\epsilon P_{max(0)} 
\end{equation}

One can easily see that upon the intensity increase the error factor becomes less and 
so the channel tends to a self-correction if the beam gets more intensive. 

\newpage

\section{Channel security}
The channel security is it's stability against the eavesdropping attacks.  Let some eavesdropping 
intruder (her name is Eve) tries to obtain the secret key, which is transfered through the channel to 
Alice and Bob. Eve can use various eavesdropping techniques, but anyway her intrusion changes the 
statistical properties of the state, which can be described in the terms of the density matrixes.

The density matrix of the TMCC-source is 

\begin{equation}
\rho_s 
= \left| \lambda \right\rangle \otimes \left\langle \lambda \right| = \frac{1}{I_0 
(2\left| \lambda \right|)}\sum\limits_{n,m = 0}^\infty {\frac{\bar {\lambda 
}^m \lambda ^n}{m!n!}\left| {mm} \right\rangle \otimes \left\langle {nn} 
\right|} 
\end{equation}

It consists of non-diagonal elements which correspond to the correlation between twin beams. 
Alice's detector reduces the source state by the Alice's mode states: 

\begin{equation}
\rho_s \longrightarrow \rho_2 = { }_1\left\langle k 
\right|\rho_s \left| k \right\rangle _1 
\end{equation}

and turns non-diagonal elements to zero. So the density matrix of the mode, which is due to be 
measured by Bob is

\begin{equation}
\label{eq:bob_rho}
\rho_B = \left\langle {\rho _s } \right\rangle _A = 
Tr_A \rho_s = \frac{1}{I_0 (2\left| \lambda \right|)}\sum\limits_{k 
= 0}^\infty {\frac{\left| \lambda \right|^{2k}}{k!^2}\left| k \right\rangle \otimes
\left\langle k \right|} = \sum\limits_{k = 0}^\infty {P_k \left| k 
\right\rangle \otimes \left\langle k \right|}
\end{equation}

When Eve starts the interception, she begins to measure one of the modes (Bob's, for example)
with her own detector, and so the density matrixes, describing the measurement results for each of the
detectors, will be

\begin{equation}
\tilde {\rho}_A = Tr_{BE} \rho_s \\
\tilde {\rho}_B = Tr_{AE} \rho_s \\
\tilde {\rho}_E = Tr_{AB} \rho_s
\end{equation}

If Eve is intercepting the Bob's mode, her intrusion will not change the density matrix for Alice, 
$\tilde {\rho}_A = \rho_A$.

We estimate the distance between the density matrixes by 
means of Hilbert-Schmidt norm as 
$D^2 = ||\rho_{(B,E)} - \tilde {\rho }_{(B,E)} ||^2$, or by means of the weak norm $d = \left| {\rho_{(B,E)} - 
\tilde {\rho }_{(B,E)} } \right|$. Since both of the matrixes are diagonal, 
the distance by the weak norm is calculated as the maximum of the 
difference absolute value. These observables may be helpful for the detection of an interception
or for estimating the successfulness of an eavesdropping. They can be obtained
using the probabilities distributions:

\begin{equation}
D^2 = \sum_{n=0}^\infty(P_n^{(orig)} - \tilde P_n^{(B,E)})^2
\end{equation}

\begin{equation}
d=max_{n=0..\infty}(|P_n^{(orig)}-\tilde {P}_n^{(B,E)}|)
\end{equation}

\subsection{Beam splitting attack}
\begin{figure}[htbp]
	\epsfbox{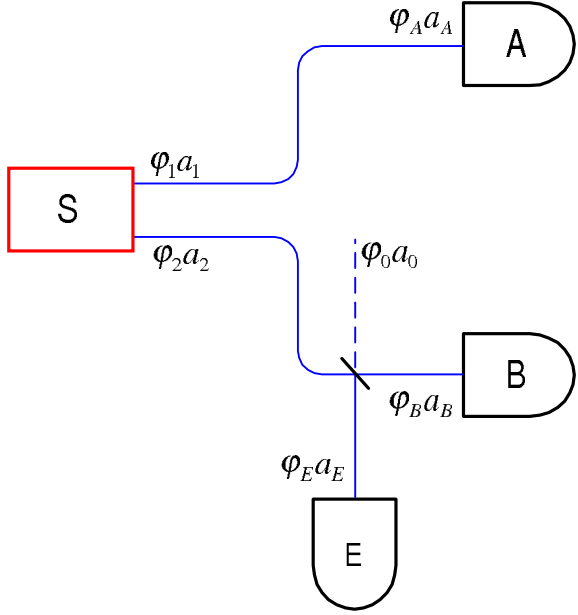}
	\caption{Eavesdropping attack on a TMCC-based quantum channel}
	\label{scheme2}
\end{figure}

The basic type of the eavesdropping attacks is the beam splitting, when Eve splits and averts a part of the beam, 
which goes to  Bob and detects its intensity by installing a detector at her side (\fref{scheme2}). 
The field amplitude of the beam splits then in some $p:q$ ratio and thus 
instead of the quantum mode we have to use the superposition 

\begin{eqnarray}
A = \varphi _A^\ast (x,t)a_A^ + + \varphi _A (x,t)a_A + \varphi _B (x,t)a_B 
+ \varphi _B^\ast (x,t)a_B^ + + \nonumber\\
+ \varphi _E (x,t) a_E  +  \varphi _E^\ast (x,t)a_E^ +
\end{eqnarray}

Obviously, $\varphi _B (x_B ,t_0 ) = 1$,$\varphi _E (x_E ,t_0 ) = 1$ and 
$\varphi _E (x_B ,t_0 ) = 0$, $\varphi _B (x_E ,t_0 ) = 0$, $\varphi _A (x_E 
,t_0 ) = 0$. The 2$^{nd}$ mode is decomposed on the basis, which consists of 
the modes coming to Bob and Eve. In order to describe the properties of 
this beam, we add a mode to the basis of Bob and Eve, which is 
orthogonal to $\varphi _2 $:

\begin{equation}
\varphi _0 = - q\varphi _B + p\varphi _E 
\end{equation}

Without the eavesdropping (and, thus, without the splitter), Eve receives 
only the $\varphi _0 $ mode, in which the laser doesn't radiate, i.e. 
$\varphi _0 = \varphi _E $ and $\varphi _2 = \varphi _B $.

The following conversion of operators corresponds to this decomposition:

\begin{equation}
a_2 = pa_B + qa_E ,
\end{equation}

\begin{equation}
a_0 = - qa_B + pa_E 
\end{equation}

Thus

\begin{equation}
\varphi _0 a_0 + \varphi _2 a_2 = \varphi _B a_B + \varphi _E a_E, 
\end{equation}

\noindent
and similarly for the hermitian-conjugate operators.

These transformations change the state (\ref{eq:tmccground}) to:

\begin{equation}
\left| \lambda \right\rangle = \frac{1}{\sqrt{I_0\left(2\left|\lambda\right|\right)}}I_0 (\lambda a_A^ + (pa_B^ + + qa_E^ + 
)\left| 0 \right\rangle 
\end{equation}

Accordingly the probability distributions of photon numbers detection change:

\begin{equation}
\tilde {P_n^B}=\frac{\left| \lambda \right|^{2n} |p|^{2n} I_n \left( 
{2\left| q \right|\left| \lambda \right|} \right)}{|q|^nn!I_0 \left( {2\left| \lambda \right|} 
\right)}
\end{equation}

\begin{equation}
\tilde {P_n^E}=\frac{\left| \lambda \right|^{2n} |q|^{2n} I_n \left( 
{2\left| p \right|\left| \lambda \right|} \right)}{|p|^nn!I_0 \left( {2\left| \lambda \right|} 
\right)}
\end{equation}

\begin{figure}[htbp]
\begin{center}
	\epsfbox{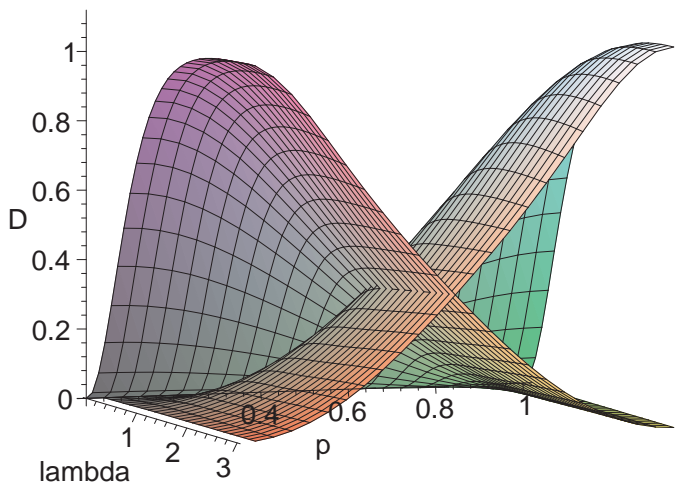}
	\caption{Distances between density matrixes for Alice-Bob and Alice-Eve pairs dependence on the intensity  of the source beam $\lambda$ and the extent $p=cos\psi$ of eavesdropping}
	\label{plot_D}
\end{center}
\end{figure}

Knowing these distributions we can estimate if the beam splitting was successfull by calculating the 
distances between the density matrixes  for Bob-Alice and Eve-Alice pairs. The dependence of these distances on
the intensity of the source beam and the extent of eavesdropping is given at the (\fref{plot_D}).

One can see that in the case of the weak intercept the results of the Bob's measurements 
almost do not change, but the eavesdropping isn't effective. If it becomes effective, Bob 
detects losses in the transmission quality and the channel gets destroyed.

\begin{figure}[htbp]
\begin{center}
	\epsfbox{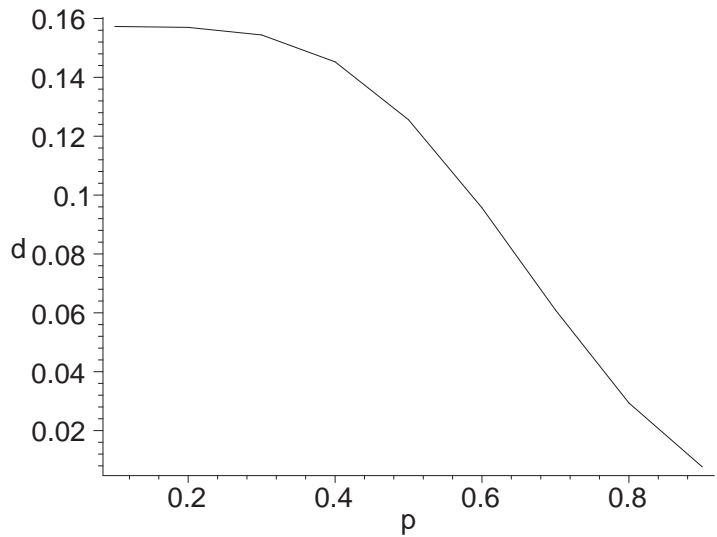}
	\caption{The dependence of the weak norm distance between actual and expected density matrixes on the extent of eavesdropping.  $\left|\lambda\right|=2$.}
	\label{plot_weak}
\end{center}
\end{figure}

Besides Bob can calculate the weak norm distance between the density matrixes for the received and 
expected states in order to check whether the beam was splitted or not. The dependence of the weak 
norm distance between actual and expected density matrixes on the extent of eavesdropping is given at
the (\fref{plot_weak}).

\subsection{State cloning attack}
Besides the basic eavesdropping, based on the beam splitting, Eve can try to 
measure the whole Bob's mode, which comes from the laser source and then 
clone the state by re-emitting the same photon number in Bob's direction 
using her own laser source. She can do it in three ways:

\begin{enumerate}
	\item Eve has a large enough number of the single-photon laser sources and she 
switches on the required number of these;
	\item She has the usual coherent laser source which can be set up to produce the 
required average number of photons;
	\item She uses the TMCC-beam source to produce the required average number of 
photons in a TMCC-state, sending one of the modes to Bob and discarding another one.
\end{enumerate}

Still the statistical properties of the cloned state will not be the same. 
For the case of the multiple single-photon sources we may assume that Bob's 
detector can distinguish between the pure n-photon state and the mixture n 
one-photon states, produced by different non-correlated sources. If Eve uses 
the usual laser beam for the state cloning, it should be noticed that such 
beam has well-defined phase, which can be identified using the interference, 
for example.

Now let's assume Eve uses the TMCC source. In order to clone the state she 
has to guess which value of the state parameter $\lambda $ corresponds to 
the exact number of photons, measured by Eve for the next incoming pulse. 
The optimal strategy for Eve to get the required value of $\lambda$ is 
the numerical solution of the state equation. Assuming Eve measured n 
photons in the Bob's mode, she tries to produce the same number of photons 
by setting her laser up to the calculated state parameter $\lambda (n)$. 
When Eve builds the cloned states, the $\lambda (n)$  is  includes an 
arbitrary phase multiplier, but it doesn't change the density matrix for Bob. This fact 
is the advantage of a TMCC-source based cloning technique. 

One mode of the cloned state is discarded and another, which will be in fact 
received and measured by Bob is averaged by the states of the discarded mode (similarly to the (\ref{eq:bob_rho}). 
Since the cloned state, under condition that Eve measured n photons, will give Bob the 
density matrix $\tilde {\rho }_B^{(n)} = \rho(\lambda 
(n))$, the full density matrix, corresponding to Bob's measurement 
results, is the mixture 

\begin{equation}
\tilde {\rho }_B = \sum\limits_{n = 0}^\infty{\tilde {\rho }_B^{(n)} P_{E,n} (\lambda )},
\end{equation}

of the cloned states for different n with weights 

\[P_{E,n} (\lambda )=\frac{\left| \lambda \right|^{2n}}{n!^2I_0 
(2\left| \lambda \right|)},\] 
which are equal to the probabilities of n photons registration by Eve. 
Finally the density matrix, corresponding to Bob's measurement has the form of  mixture of  k-photon states
\begin{equation}
\tilde {\rho }_B=\sum\limits_{k = 0}^\infty {\tilde {P}_k 
\left| k \right\rangle \otimes \left\langle k \right|}
\end{equation}

with probabilities 
\begin{equation}
\tilde {P}_k = 
\sum\limits_{n = 0}^\infty {
\frac{\left| \lambda \right|^{2n}}{n!^2I_0 
(2\left| \lambda \right|)}
\frac{\left| {\lambda (n)} \right|^{2k}}
{k!^2I_0(2\left| {\lambda (n)} \right|)}} .
\end{equation}

\begin{figure}[htbp]
	\epsfbox{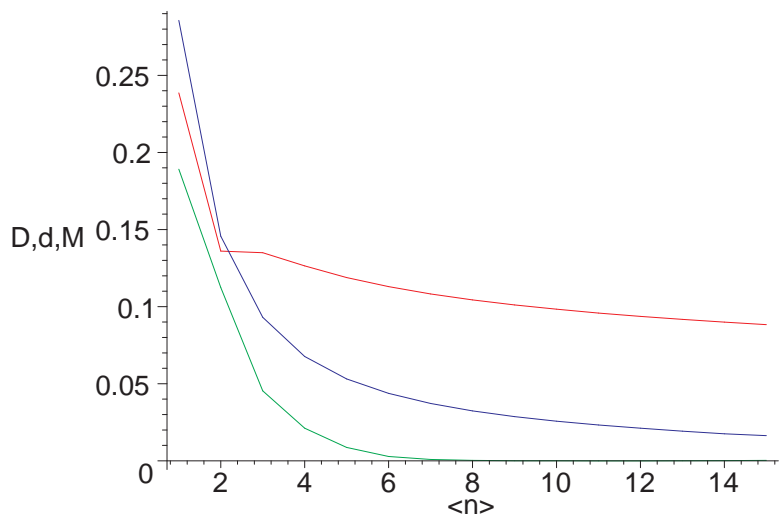}
	\caption{Mandel parameter of the cloned state (green line) and the density matrixes distances between 
	the cloned and expected states by Hilbert-Schmidt (blue line) and weak (red line) norms.}
	\label{plot_cloning}
\end{figure}

The change of the probabilities distribution of the state can be easily 
estimated by Bob by comparison of the Mandel parameter of the cloned beam to 
the expected value. The successfulness of cloning can be estimated by
calculating the distance between actual and expected density matrixes. 
The corresponding graphs, showing the dependence of 
these values on the intensity of a source beam are given at the (\fref{plot_cloning}). 

One can easily see that the state cloning attempts change the Mandel parameter 
and that density matrix of the cloned state differs from 
the density matrix of the original state even in the case of the optimal 
cloning strategy. Thus a cloning attack on a TMCC-based secure channel can 
be detected and so isn't effective.

\newpage

\section{Conclusions}

Correlated coherent states of the two-mode laser beam (TMCC states) show 
interesting properties, which can be used, in particular, for the tasks of 
the quantum communication and cryptography. 

The TMCC-beams can be identified due to the special form of the registration 
probabilities distribution for different photon numbers in the corresponding beam 
and the dependence of the dispersion on the mean photon numbers value.

On the one hand, each of the 
modes looks like a flow of the independent photons rather then a coherent 
beam, since mean values of the operators, which are linear in field, are 
equal to 0 for each mode separately. 

On the other hand, the strong 
correlation between the results of measurements for each of the modes takes 
place. This correlation shows itself in the fact that in each of the modes numbers 
of photons are the same and even the shot noise shows itself equally in the both 
modes. This enables the use of the TMCC state as the generator and carrier 
of random keys in a quantum channel which is stable against the eavesdropping \cite{tmcc}.

Thus, the TMCC-laser generates and transmits exactly the 2 copies of a 
random key. Unlike the single or two-photon schemes, which require large numbers of 
transmission reiterations to obtain the statistically significant 
results, the TMCC beam can be intensive enough to make each single measurement statistically 
significant and thus to use single impulse for each 
piece of information, and remain cryptographically steady. This allows to essentially increase 
the effective data transfer rate and distance. Analysis of the noise influence on
the channel properties shows that the channel tends to a self-correction upon the beam intensity increase.

The TMCC-based channel turns to be stable against the beam splitting and state cloning eavesdropping, which
are either non-effective or significant and easy to be detected.


\Bibliography{99}
\bibitem{qc1} Nicolas Gisin, Gregoire Ribordy, Wolfgang Tittel, Hugo Zbinden. Quantum Cryptography. Preprint: quant-ph/0101098
\bibitem{qc2} Matthias Christandl, Renato Renner, Artur Ekert.  A Generic Security Proof for Quantum Key Distribution. Preprint: quant-ph/0402131
\bibitem{qc3} Nicolas Gisin, Nicolas Brunner. Quantum cryptography with and without entanglement. Preprint: quant-ph/0312011
\bibitem{per} Perelomov A.V. - Generalized coherent states, M. 1982 (in Russian)  
\bibitem{ent1} Wolfgang Tittel, Gregor Weihs. Photonic Entanglement for Fundamental Tests and Quantum Communication. quant-ph/0107156
\bibitem{entprot} A. Ekert, Phys. Rev. Lett. 67, 661 (1991) 
entprotexp D. S. Naik et al., Phys. Rev. Lett. 84, 4732 (2000)
\bibitem{single1} C. H. Bennett, Phys. Rev. Lett. 68, 3121 (1992)
\bibitem{single2} C. K. Hong and L. Mandel, Phys. Rev. Lett. 56, 58 (1986)
\bibitem{four} C. H. Bennett and G. Brassard , "Quantum cryptography: public key distribution and coin tossing", Int . conf. Comput ers, Syst ems \& Signal Processing, Bangalore, India, 1984, 175- 179.
\bibitem{tmcc} Constantin V. Usenko and Vladyslav C. Usenko. arxiv.org/quant-ph/0403112 (accepted to Journal of Russian Laser Research)
\bibitem{tmcc2} Constantin V. Usenko and Vladyslav C. Usenko. arxiv.org/quant-ph/0404131
\bibitem{fourexp} T. Jennewein et al., Phys. Rev. Lett. 84, 4729 (2000)
\bibitem{polar} A.C. Funk, M.G. Raymer. Quantum key distribution using non-classical photon number correlations in macroscopic light pulses. quant-ph/0109071
\bibitem{similar1} Yun Zhang, Katsuyuki Kasai, Kazuhiro Hayasaka. Quantum channel using photon number correlated twin beams. quant-ph/0401033, Optics, Express 11, 3592 (2003)
\bibitem{similar2} L. A. Wu, H. J. Kimble, J. L. Hall, and H. F. Wu, "Generation of squeezed states by parametric down conversion," Phys. Rev. Lett. 57, 2520-2524 (1986). 
\bibitem{similar3} H. Wang, Y. Zhang, Q. Pan, H. Su, A. Porzio, C. D. Xie, and K. C. Peng, "Experimental realization of a quantum measurement for intensity difference fluctuation using a beam splitter," Phys. Rev. Lett. 82, 1414-1417 (1999). 

\endbib
\end{document}